\documentclass[aps,prd,12pt,preprint,tightenlines,superscriptaddress,
amsfonts,amssymb,amsmath,byrevtex,showpacs,nofootinbib]{revtex4}
\usepackage{graphics}
\usepackage{epsfig,subfigure}
\usepackage{amsfonts}
\newcounter{mnotecount}[section]
\renewcommand{\themnotecount}{\thesection.\arabic{mnotecount}}
\newcommand{\mnote}[1]
{\protect{\stepcounter{mnotecount}}$^{\mbox{\footnotesize  $
      \bullet$\themnotecount}}$ \marginpar{\raggedright\tiny
    $\!\!\!\!\!\!\,\bullet$\themnotecount: #1} }

\begin{document}
\newcommand{\dR}{\mathbb R}
\newcommand{\dC}{\mathbb C}
\newcommand{\dZ}{\mathbb Z}
\newcommand{\id}{\mathbb I}
\newcommand{\dT}{\mathbb T}

\title{Bianchi I model in terms of nonstandard loop quantum cosmology: Quantum dynamics.}

\author{ Przemys{\l}aw Ma{\l}kiewicz$^\dag$,
W{\l}odzimierz Piechocki$^\ddag$, and Piotr Dzier\.{z}ak$^\S$
\\ Theoretical Physics Department, Institute for Nuclear Studies
\\ Ho\.{z}a 69, 00-681 Warsaw, Poland;
\\ $^\dag$pmalk@fuw.edu.pl, $^\ddag$piech@fuw.edu.pl, $^\S$pdzi@fuw.edu.pl }

\date{\today}

\begin{abstract}
We analyze the quantum Bianchi I model in the setting of the
nonstandard loop quantum cosmology. Elementary observables are
used to quantize the volume operator. The spectrum of the volume
operator is bounded from below and discrete. The discreteness may
imply a foamy structure of spacetime at semiclassical level.  The
results are described in terms of a free parameter specifying loop
geometry to be determined in astro-cosmo observations. An
evolution of the quantum model is generated by the so-called true
Hamiltonian, which enables an introduction of a time parameter
valued in the set of all real numbers.

\end{abstract}

 \maketitle

\section{Introduction}

The great challenge is quantization of the
Belinskii-Khalatnikov-Lifshitz (BKL) theory \cite{BKL1,BKL2,BKL3}.
It is a generic solution of General Relativity (GR) that does not
rely on any symmetry conditions. It presents an evolution of the
universe near the space-like cosmological singularity (CS) with
diverging gravitational and matter fields invariants. The singular
solution may be applied both to the future singularity (Big
Crunch) and past singularity (Big Bang). The BKL scheme also
appears in the low energy limit of superstring models, where it is
linked to the hyperbolic Kac-Moody algebras \cite{Damour:2002et}.

The dynamics of the BKL model, close to the singularity, may be
approximated by  the two key ingredients: (i) Kasner-type
evolutions well approximated by the Bianchi I model, and (ii)
spike-type transitions described by the Bianchi II model. The
former case occurs when  time derivatives in the equations of
motion are important, whereas the latter case appears when
space-like gradients play the crucial role. The general BKL type
evolution, near the cosmological singularity, consists of a
sequence of epochs (i) and (ii), which may lead to the oscillatory
and chaotic type dynamics.

Present paper concerns quantization of the Bianchi I model  with
massless scalar field. It is a companion paper to our recent paper
\cite{Dzierzak:2009dj}, presenting classical dynamics of the
Bianchi I model in terms of  loop geometry.

Through the paper we apply the reduced phase space, RPS,
quantization method developed by us recently
\cite{Dzierzak:2008dy,Dzierzak:2009ip,Malkiewicz:2009zd,Malkiewicz:2009qv,Mielczarek:2010rq}.
It is an alterative method to the Dirac quantization recently
applied to the quantization of the Bianchi I model
\cite{Bojowald:2003md,Chiou:2006qq,Chiou:2007mg,Szulc:2008ar,MartinBenito:2009qu,Ashtekar:2009vc}.

In section II, we recall some elements of the classical formalism
\cite{Dzierzak:2009dj} for self-consistency. We redefine an
evolution parameter and elementary observables, and introduce the
true Hamiltonian. Section III is devoted to the quantization of
the classical model. Compound observables are quantized in terms
of elementary observables. We present solution to the eigenvalue
problem for the volume operators. Examination of an evolution of
the quantum system completes this section. We conclude in the last
section.  Unitarily non-equivalent representations of the volume
operators are briefly  discussed  in  the appendix A. We make
comments on the operators ordering problem in the appendix B.

\section{Preparations to quantization}

The Bianchi I model with massless scalar field  is described by
the line element
\begin{equation}\label{bia}
ds^2= -N^2\,dt^2 + \sum_{i=1}^{3} a_i^2(t)\,dx_i^2,
\end{equation}
where
\begin{equation}\label{bbb}
a_i(\tau)=
a_i(0)\,\bigg(\frac{\tau}{\tau_0}\bigg)^{\textrm{k}_i},~~~~d\tau=
N\,dt,~~~~\sum_{i=1}^{3} \textrm{k}_\textrm{i}= 1 = \sum_{i=1}^{3}
\textrm{k}_\textrm{i}^2 + \textrm{k}_{\phi}^2,
\end{equation}
and where $\,\textrm{k}_{\phi}$ describes matter field density
($\textrm{k}_{\phi}= 0$ corresponds to the Kasner model).  For a
clear exposition of the {\it classical} singularity aspects of the
Bianchi I model, in terms of the loop geometry, we recommend
\cite{Chiou:2006qq}.

\subsection{Hamiltonian constraint}

The gravitational part of the classical Hamiltonian, for the
Bianchi I model with massless scalar field, reads
\cite{Dzierzak:2009dj}
\begin{equation}\label{hamG}
H_g = - \gamma^{-2} \int_{\mathcal V} d^3 x ~N
e^{-1}\varepsilon_{ijk}
 E^{aj}E^{bk} F^i_{ab}\, ,
\end{equation}
where  $\gamma$ is the Barbero-Immirzi parameter, $\mathcal
V\subset \Sigma$ is the fiducial volume, $\Sigma$ is spacelike
hypersurface,  $N$ denotes the lapse function, $\varepsilon_{ijk}$
is the alternating tensor, $E^a_i $ is a densitized  vector field,
$e:=\sqrt{|\det E|}$, and where $F^i_{ab}$ is the curvature of an
$SU(2)$ connection $A^i_a$.

The resolution of the singularity, obtained within LQC, is based
on rewriting  the curvature $F^k_{ab}$ in terms of the holonomy
around a loop by making use of the formula \cite{Dzierzak:2009dj}
\begin{equation}\label{cur}
F^k_{ab}= -2~\lim_{Ar\,\Box_{ij}\,\rightarrow \,0}
Tr\;\Big(\frac{h_{\Box_{ij}}-1}{Ar\,\Box_{ij}}\Big)\;{\tau^k}\;
^o\omega^i_a  \; ^o\omega^j_a ,
\end{equation}
where
\begin{equation}\label{box}
h_{\Box_{ij}} = h^{(\mu_i)}_i h^{(\mu_j)}_j (h^{(\mu_i)}_i)^{-1}
(h^{(\mu_j)}_j)^{-1}
\end{equation}
is the holonomy of the gravitational connection around the square
loop $\Box_{ij}$. The loop is taken over a face of an elementary
cell, each of whose sides has length $\mu_j L_j$  with respect to
the flat {\it fiducial} metric $^o q_{ab}:= \delta_{ij}\, ^o
\omega^i_a\, ^o \omega^j_a $; the fiducial triad $^o e^a_k$ and
cotriad $^o \omega^k_a$ satisfy $^o \omega^i_a\,^o e^a_j =
\delta^i_j$; $~Ar\,\Box_{ij}$ denotes the area of the square
$\Box_{ij}$; $L_1 L_2 L_3= V_0$, where $V_0 = \int_{\mathcal V}
\sqrt{^o q} d^3 x$ is the fiducial volume of $\mathcal V$ with
respect to the fiducial metric.

The holonomy in the fundamental, $j=1/2$, representation of
$SU(2)$ reads
\begin{equation}\label{hol}
h^{(\mu_i)}_i  =\cos (\mu_i c_i/2)\;\id + 2\,\sin (\mu_i
c_i/2)\;\tau_i,
\end{equation}
where $\tau_i = -i \sigma_i/2\;$ ($\sigma_i$ are the Pauli spin
matrices). The  connection $A^k_a$ and the density weighted triad
$E^a_k$ are determined by the conjugate variables $c_k$ and $p_k$
as follows
\begin{equation}
A^i_a = \,c_i\,{L_i}^{-1}\,^o\omega^i_a, ~~~~E^a_i = \,p_i\,{L_j}^{-1}
\,{L_k}^{-1}\,^oe^a_i ,
\end{equation}
where
\begin{equation}\label{pici}
c_i = \gamma\,\dot{a_i}\,L_i,~~~~ |p_i| = a_j\,a_k\,L_j\,L_k ,
\end{equation}
( the dot over $a_i$ denotes derivative with respect to the
cosmological time), and where
\begin{equation}
\{c_i, p_j\}= 8\pi G \gamma \delta_{ij}
\end{equation}

Making use of (\ref{hamG}) and (\ref{cur}) leads to $H_g$ in the
form \cite{Dzierzak:2009dj}
\begin{equation}\label{hamR}
    H_g = \lim_{\mu_1,\mu_2,\mu_3\rightarrow \,0}\; H^{(\mu_1\,\mu_2\,\mu_3)}_g ,
\end{equation}
where
\begin{equation}\label{hamL}
H^{(\mu_1\,\mu_2\,\mu_3)}_g = - \frac{\text{sign}(p_1p_2p_3)}{2\pi
G \gamma^3 \mu_1\mu_2\mu_3} \sum_{ijk}\,N\, \varepsilon^{ijk}\, Tr
\Big(h^{(\mu_i)}_i h^{(\mu_j)}_j (h^{(\mu_i)}_i)^{-1}
(h^{(\mu_j)}_j)^{-1} h_k^{(\mu_k)}\{(h_k^{(\mu_k)})^{-1},V\}\Big),
\end{equation}
and where $V= a_1a_2a_3 V_0$ is the volume of the elementary cell
$\mathcal{V}$ with respect to the {\it physical} metric $q_{ab}=
a_1a_2a_3\,^o q_{ab}$.

The total Hamiltonian for Bianchi I universe with a massless
scalar field, $\phi$, reads
\begin{equation}\label{ham}
   H = H_g + H_\phi \approx 0,
\end{equation}
where $H_g$ is defined by (\ref{hamR}). The Hamiltonian of the
scalar field  reads $H_\phi = N\,p^2_\phi |p_1 p_2
p_3|^{-\frac{1}{2}}/2$, where $\phi$ and $p_\phi$ are the
elementary variables satisfying $\{\phi,p_\phi\} = 1$. The
relation $H \approx 0$ defines the constraint on phase space of
considered gravitational system.

Making use of (\ref{hol}) we calculate  (\ref{hamL}) and get the
{\it modified}  total Hamiltonian $H^{(\lambda)}_g$ corresponding
to (\ref{ham}) in the form
\begin{equation}\label{regH}
  H^{(\lambda)}/N= -\frac{1}{8\pi G \gamma^2}\;\frac{\text{sgn}(p_1p_2p_3)}
  {\mu_1\mu_2\mu_3}\bigg[\sin(c_1 \mu_1)\sin(c_2
\mu_2)\,\mu_3\;\textrm{sgn}(p_3)\sqrt{\frac{|p_1p_2|}{|p_3|}} +
\textrm{cyclic}\bigg] + \frac{p_{\phi}^2}{2\,V},
\end{equation}
where $V = \sqrt{|p_1 p_2 p_3|}$. In what follows we assume that
\begin{equation}\label{re1}
\mu_k:= \sqrt{\frac{1}{|p_k|}}\,\lambda ,
\end{equation}
where $\lambda$ is  a free {\it parameter} of our model.

The choice (\ref{re1}) for $\mu_k$ leads, in the Dirac
quantization \cite{Chiou:2007mg,Szulc:2008ar,Ashtekar:2009vc}, to
the dependance of dynamics on $\mathcal{V}\subset\Sigma$. In what
follows, we specialize our considerations to the case
$\mathcal{V}= \Sigma = \dT^3$. In such a case the volume {\it
observable},  $V = \int_{\mathcal V} \sqrt{q}\, d^3 x$ (where $g$
denotes the determinant of the physical metric $q_{ab}$ on
$\Sigma$), characterizes  the entire space part of the universe.
Thus, $\mathcal{V}$ is chosen unambiguously  and $V$ is physical.
In the case when one considers the Bianchi I model with the
$\dR^3$ topology, the volume $\mathcal{V} \subset \Sigma = \dR^3$
is only an auxiliary {\it tool} devoid of any physical
meaning\footnote{It is tempting to introduce the notion of a {\it
local} volume,  i.e. a sort of `density' of the physical volume
operator $V$. It appears that the local volume may be defined in
the case of {\it any} topology of space, however it would depend
on the choice of coordinates on $\Sigma$. Similar notion has been
used by Martin Bojowald (see, Appendix in \cite{Bojowald:2007ra}),
while considering lattice refining in loop quantum cosmology.}.

The present paper is a quantum version of our recent paper
\cite{Dzierzak:2009dj}, where we consider the Bianchi I model with
the $\dT^3$ topology. The aim of both our papers is presenting a
quantum Bianchi I model in terms of the {\it nonstandard} LQC,
which is an alternative to the {\it standard} LQC results
\cite{Szulc:2008ar,MartinBenito:2009qu} (with $\mathcal{V}= \dT^3$
and the choice (\ref{re1})). The results obtained within these two
methods are similar. Detailed comparison is beyond the scope of
the present paper, but will be presented elsewhere after we
complete quantization of the Bianchi II model \cite{MGWP}.

Let us analyze the dependance of our results on the choice of
coordinates in $\Sigma$, and consequently on the choice of the
fiducial volume $V_0 = L_1 L_2 L_3$. It is clear that $L_k$ is a
{\it coordinate} length, whereas $a_k L_k$ is the {\it physical}
one. The latter is invariant with respect to a change of the
system of coordinates so we have $a_k L_k = a_k^\prime
L_k^\prime$, while $L_k\rightarrow L_k^\prime$.  Since, due to
(\ref{pici}), we have
\begin{equation}\label{ch6}
c_i = \gamma\frac{\dot{a_i}}{a_i}\,a_i L_i,~~~~ |p_i| = a_j
L_j\,a_k L_k ,
\end{equation}
the canonical variables $c_i$ and $p_i$  do not depend on the
choice of the coordinate system. Thus, the destination variables,
defined by Eq. (\ref{haam}), share this property too. The holonomy
variable $h^{(\mu_k)}_k$ depends on $\mu_k c_k$, and with our
choice (\ref{re1}) leads to
\begin{equation}\label{cor3}
\mu_k c_k = \lambda \frac{c_k}{\sqrt{|p_k|}} = \lambda^\prime
\frac{c_k^\prime}{\sqrt{|p_k^\prime|}} = \mu_k^\prime c_k^\prime ,
\end{equation}
which proves that the holonomy variable does not depend on the
choice of coordinates\footnote{We have $\lambda = \lambda^\prime$
since $\lambda$ is a {\it physical} length.}. The flux variable is
$p_k$ so it does not depend on the choice of coordinates either.
Since holonomy and flux are basic variables of the formalism, our
final results do not depend on the choice of coordinates (the
choice of $V_0$). This is why the variables $\beta_k$ and $v_k$
share  this property as well. Our results do depend on
$\mathcal{V}$, but it is correct since $\mathcal{V} = \dT^3$ is
the whole space.

We wish to emphasize that (\ref{regH}) is not an {\it effective}
Hamiltonian for quantum dynamics \cite{Chiou:2007mg}, but a {\it
classical} Hamiltonian {\it modified} by  approximating the
curvature $F^k_{ab}$ by holonomy of connection around a loop with
{\it finite} length. Our approach is quite different from the
so-called polymerization method where the replacement $c
\rightarrow \sin(c \mu )/\mu$ in the Hamiltonian is treated as
some kind of  an effective {\it quantization}. Our method has been
presented with all details and compared with the Dirac
quantization method  in
\cite{Dzierzak:2008dy,Dzierzak:2009ip,Malkiewicz:2009qv,Mielczarek:2010rq}.
For an extended {\it motivation} of our approach we recommend an
appendix of \cite{Malkiewicz:2009qv}.

In the gauge $N= V = \sqrt{|p_1 p_2 p_3|}$, the Hamiltonian
modified by loop geometry reads
\begin{equation}\label{re33}
H^{(\lambda)}= -\frac{1}{8\pi
G\gamma^2\lambda^2}\;\bigg[\text{sgn}(p_1p_2)|p_1p_2|^{3/2}\sin\big(\lambda
\frac{c_1}{\sqrt{|p_1|}}\big)\sin\big(\lambda
\frac{c_2}{\sqrt{|p_2|}}\big) +\textrm{cyclic} \bigg] +
\frac{p_{\phi}^2}{2}.
\end{equation}
Since we consider the {\it relative} dynamics
\cite{Dzierzak:2009ip,Dzierzak:2009dj} our results, in what
follows, are gauge independent.

Equation (\ref{re33}) corresponds to the effective quantization of
the standard LQC. In the nonstandard LQC Eq. (\ref{re33}) is
treated as the constraint, which is to be imposed into the {\it
classical} dynamics. In the standard LQC one implements this
constraint into an operator constraint defining {\it quantum}
dynamics (kernel of this operator is used to find the physical
Hilbert space). Thus, in the reduced phase space quantization
(nonstandard LQC) there is no quantum Hamiltonian constraint,
contrary to the Dirac quantization, i.e. the standard LQC (which
is motivated from LQG). In both cases one {\it modifies}, to some
extent, gravity theory: already at the classical level in the
reduced phase space quantization, only at the quantum level in the
Dirac quantization.

The Poisson bracket is defined to be
\begin{equation}\label{re2}
    \{\cdot,\cdot\}:= 8\pi G\gamma\;\sum_{k=1}^3\bigg[ \frac{\partial \cdot}
    {\partial c_k} \frac{\partial \cdot}{\partial p_k} -
     \frac{\partial \cdot}{\partial p_k} \frac{\partial \cdot}{\partial c_k}\bigg] +
     \frac{\partial \cdot}{\partial \phi} \frac{\partial \cdot}{\partial p_\phi} -
     \frac{\partial \cdot}{\partial p_\phi} \frac{\partial \cdot}{\partial
     \phi} ,
\end{equation}
where $(c_1,c_2,c_3,p_1,p_2,p_3,\phi,p_\phi)$ are canonical
variables.  The dynamics of  a function $\xi$ on a phase space is
defined by
\begin{equation}\label{dyn}
    \dot{\xi} := \{\xi,H^{(\lambda)}\},~~~~~~\xi \in
    \{c_1,c_2,c_3,p_1,p_2,p_3,\phi,p_\phi\}.
\end{equation}
The dynamics is defined by the solutions to (\ref{dyn}) satisfying
the constraint $H^{(\lambda)}\approx 0$. The solutions of
(\ref{dyn}) ignoring the constraint are {\it nonphysical}.

In what follows we shift from $(c_k, p_k)$ to another canonical
variables $(v_k,\beta_k)$
\begin{equation}\label{haam}
\beta_k := \frac{c_k}{\sqrt{|p_k|}},~~~~v_k:= |p_k|^{3/2},
\end{equation}
(where $k = 1,2,3$) since they are proper variables to examine the
singularity aspects of our model \cite{Dzierzak:2009dj}. In this
paper we restrict our considerations to $v_k \geq 0$, since we
wish to ascribe to it a directional {\it volume} observable.

\subsection{Correspondence with  FRW observables}

We redefine the original elementary Bianchi observables that has
been found in \cite{Dzierzak:2009dj} as follows
\begin{equation}\label{O2}
\textrm{O}_{i}=
\frac{1}{3\kappa\gamma}\,\frac{v_i\sin(\lambda\beta_i)}{\lambda},
\end{equation}
and
\begin{equation}\label{A2}
\textrm{A}_i=
\frac{1}{3\kappa}\ln{\bigg(\frac{\big|\tan\big(\frac{\lambda\beta_i}{2}
\big)\big|}{\frac{\lambda}{2}}\bigg)} + \frac{3}{2\sqrt{3}}
\frac{\textrm{sgn}(p_{\phi})\big(\textrm{O}_j+\textrm{O}_k\big)\,
\phi}{\sqrt{\textrm{O}_1\textrm{O}_2 + \textrm{O}_1\textrm{O}_3 +
\textrm{O}_2\textrm{O}_3}},
\end{equation}
where $\kappa^2 := 4\pi G/3$. Dropping subscripts leads to the
elementary observables of the FRW model found in
\cite{Malkiewicz:2009qv}. One may verify that the algebra of
redefined observables is
\begin{equation}\label{alg prim}
\{\textrm{O}_{i},\textrm{O}_{j}\}=
0,~~~~\{\textrm{A}_{i},\textrm{O}_{j}\}=
\delta_{ij},~~~~\{\textrm{A}_{i},\textrm{A}_{j}\} = 0 .
\end{equation}

Our main concern is quantization of the {\it volume} observable
defined as follows \cite{Dzierzak:2009dj}
\begin{equation}\label{vol 1}
V = (v_1 v_2 v_3)^{1/3},
\end{equation}
where
\begin{equation}\label{vi1}
v_i =
3\kappa\gamma\lambda|\textrm{O}_{i}|\,\cosh\bigg(\frac{3\sqrt{\pi
G}\,\big(\textrm{O}_{j}+\textrm{O}_{k}
\big)\,\phi}{\sqrt{\textrm{O}_{1}\textrm{O}_{2} +
\textrm{O}_{1}\textrm{O}_{3} + \textrm{O}_{2}\textrm{O}_{3}}} -
3\kappa\textrm{A}_{i}\bigg) .
\end{equation}
In an `isotropic' case (i=j=k) we get the expression for the
volume observable of the FRW model \cite{Malkiewicz:2009qv}.

\subsection{Redefinitions of  evolution parameter}

Since the observables $\textrm{O}_\textrm{i}$ are constants of
motion in $\phi$ and $\phi \in \dR$, it is possible to make the
following redefinition of an evolution parameter
\begin{equation}\label{phip}
\varphi:= \frac{\sqrt{3
}\,\,\phi}{2\,\sqrt{\textrm{O}_\textrm{1}\textrm{O}_\textrm{2} +
\textrm{O}_\textrm{1}\textrm{O}_\textrm{3} +
\textrm{O}_\textrm{2}\textrm{O}_\textrm{3}}}
\end{equation}
so we have
\begin{equation}\label{vi2}
v_i= 3\kappa\gamma\lambda|\textrm{O}_i|\,\cosh
3\kappa\big((\textrm{O}_j+\textrm{O}_k )\,\varphi -
\textrm{A}_i\big),
\end{equation}
which simplifies further considerations.

\subsection{Redefinitions of elementary observables}

One can make the following redefinitions
\begin{equation}\label{redef}
\mathcal{A}_i := \textrm{A}_i - (\textrm{O}_j +
\textrm{O}_k)\,\varphi.
\end{equation}
Thus, the directional volume  (\ref{vi2})  becomes
\begin{equation}\label{vi5}
v_i := |w_i|,~~~~~w_i= 3\kappa\gamma\lambda O_i\,\cosh (3 \kappa
\mathcal{A}_i).
\end{equation}

The algebra of observables reads
\begin{equation}\label{alc}
\{\textrm{O}_i,\textrm{O}_j\}=
0,~~~~\{\mathcal{A}_i,\textrm{O}_j\}=
\delta_{ij},~~~~\{\mathcal{A}_i,\mathcal{A}_j\}= 0 ,
\end{equation}
where the Poisson bracket  is defined to be
\begin{equation}\label{po}
\{\cdot,\cdot\}:= \sum_{k=1}^3\Big(\frac{\partial\cdot}{\partial
\mathcal{A}_k} \frac{\partial\cdot}{\partial O_k} -
\frac{\partial\cdot}{\partial O_k} \frac{\partial\cdot}{\partial
\mathcal{A}_k}\Big).
\end{equation}

\subsection{Structure of phase space}

All considerations carried out in the previous section have been
done under the assumption that the observables $\textrm{O}_1$,
$\textrm{O}_2$ and $\textrm{O}_3$ have no restrictions.  The
inspection of (\ref{A2}), (\ref{vi1}) and (\ref{redef}) shows that
the domain of definition of the elementary observables reads
\begin{equation}\label{dom}
    D := \{(\mathcal{A}_k, \textrm{O}_k)\,|\,\mathcal{A}_k \in
    \dR,~~ \textrm{O}_1 \textrm{O}_2 + \textrm{O}_1\textrm{O}_3 +
    \textrm{O}_2 \textrm{O}_3 > 0\},
\end{equation}
where $k=1,2,3$. The restriction $\textrm{O}_1 \textrm{O}_2 +
\textrm{O}_1\textrm{O}_3 + \textrm{O}_2 \textrm{O}_3 > 0$ is a
consequence of the Hamiltonian constraint (see,
\cite{Dzierzak:2009dj} for more details).

In what follows we consider two cases:
\begin{enumerate}
    \item Kasner-unlike dynamics: (a) $\textrm{O}_i>0$, $\textrm{O}_j>0$,
    $\textrm{O}_k>0$,
    which describes all three directions expanding (b) $\textrm{O}_i<0$,
    $\textrm{O}_j<0$,
    $\textrm{O}_k<0$, with all directions shrinking.
    \item Kasner-like dynamics: (a) $\textrm{O}_i>0$, $\textrm{O}_j>0$,
    $\textrm{O}_k<0$, which describes two directions expanding and one
    direction shrinking; (b) $\textrm{O}_i <0$, $\textrm{O}_j <0$,
    $\textrm{O}_k >0$, with two directions shrinking and one
    expanding.
\end{enumerate}
This classification presents all possible nontrivial cases. Our
terminology fits the one used in \cite{Chiou:2007mg} due to the
relation $\,\textrm{O}_i= 6 \kappa k_i K,~~(0<K=const)$, where
constants $k_i$ are defined by (\ref{bbb}).

\subsection{True Hamiltonian}

Now, we define a generator of an evolution called a true
Hamiltonian $\mathbb{H}$. Making use of (\ref{redef}), and
$\textrm{O}_i = const$ (see \cite{Dzierzak:2009dj}), we get
\begin{equation}\label{true}
\{\mathcal{A}_i,\mathbb{H}\}:=\frac{d\mathcal{A}_i}{d\varphi} =-
(\textrm{O}_j + \textrm{O}_k),~~~~
\{\textrm{O}_i,\mathbb{H}\}:=\frac{d\textrm{O}_i}{d\varphi} =0 .
\end{equation}
The solution to (\ref{true}) is easily found to be
\begin{equation}\label{Htrue}
\mathbb{H}=\textrm{O}_\textrm{1}\textrm{O}_\textrm{2} +
\textrm{O}_\textrm{1}\textrm{O}_\textrm{3} +
\textrm{O}_\textrm{2}\textrm{O}_\textrm{3}.
\end{equation}
The true Hamiltonian is defined on the {\it reduced} phase space
which is devoid of constraints.

\section{Quantization}

\subsection{Representation of elementary observables}

We use the Schr\"{o}dinger representation for the algebra
(\ref{alc}) defined as
\begin{equation}\label{osy}
\textrm{O}_k\rightarrow\widehat{\textrm{O}}_k\,f_k(x_k):=
\frac{\hbar}{i}\,\frac{d}{dx_k}\,f_k(x_k),~~~\mathcal{A}_k\rightarrow
\widehat{\mathcal{A}}_k\,f_k(x_k):=x_k\, f_k(x_k),~~~k=1,2,3.
\end{equation}

One may verify that
\begin{equation}\label{almod}
[\widehat{\textrm{O}}_i,\widehat{\textrm{O}}_j]=
0,~~~~[\widehat{\mathcal{A}}_i,\widehat{\mathcal{A}}_j]=0,~~~~
[\widehat{\mathcal{A}}_i,\widehat{\textrm{O}}_j]=
i\hbar \,\delta_{ij}.
\end{equation}
The representation is defined formally on some dense subspaces of
a Hilbert space to be specified later.

\subsection{Kasner-unlike case}

The condition $\textrm{O}_1 \textrm{O}_2 +
\textrm{O}_1\textrm{O}_3 + \textrm{O}_2 \textrm{O}_3 > 0$ is
automatically  satisfied in this case, because $\textrm{O}_1,
\textrm{O}_2$ and $\textrm{O}_3$ are of the same sign. To be
specific, let us consider (1a); the case (1b) can be done by
analogy.

Let us quantize the directional volumes by means of $w_i$ defined
in (\ref{vi5}). A standard procedure gives\footnote{In what
follows we drop subscripts of observables to simplify notation.}
\begin{eqnarray}\label{qdir}
\hat{w}:=
    \frac{3\kappa\gamma\lambda}{2}\,\bigg(
    \widehat{\textrm{O}}\,\cosh \big(3\kappa\widehat{\mathcal{A}}\big)
    + \cosh
    \big(3\kappa\widehat{\mathcal{A}}\big)\;\widehat{\textrm{O}}\bigg)=
    -\frac{3ia}{2}\Big(2\cosh(bx)\frac{d}{dx}+b\sinh(bx)\Big),
\end{eqnarray}
where  $a:=\kappa\gamma\lambda\hbar\,$ and $b:=3\kappa$, and where
we have used the  representation  for the elementary observables
defined by (\ref{osy}).

In what follows we solve the eigenvalue problem for the operator
$\hat{w}$ and identify its domain of self-adjointness.

Let us consider the invertible mapping $L^2(\mathbb{R},dx)\ni \psi
\rightarrow \tilde{U}\psi=: f \in L^2(\mathbb{I},dy)$ defined by
\begin{equation}\label{interval}
 \tilde{U}\psi (x) :=\frac{\psi(\ln|\textrm{tg}^{1/b}(\frac{by}{2})|)}{\sin^{1/2}(by)}
 =:  f(y),~~x\in
    \mathbb{R},~~y\in \mathbb{I} := (0,\pi/b).
\end{equation}
We have
\begin{eqnarray}
\nonumber
    \langle\psi|\psi\rangle=\int_{-\infty}^{\infty}\overline{\psi}\psi
    ~dx\\ \nonumber
    =\int_0^{\frac{\pi}{b}}\overline{\psi}(\ln|\textrm{tg}^{1/b}(\frac{by}
    {2})|)\psi(\ln|\textrm{tg}^{1/b}(\frac{by}{2})|)d\big(\ln|\textrm{tg}^{1/b}
    (\frac{by}{2})|)\big)\\
    \nonumber
    =\int_0^{\frac{\pi}{b}}\overline{\psi}(\ln|\textrm{tg}^{1/b}(\frac{by}{2})|)
    \psi(\ln|\textrm{tg}^{1/b}(\frac{by}{2})|)\frac{dy}{\sin(by)}\\ \label{scalarproduct}
=\int_0^{\frac{\pi}{b}}\overline{\frac{\psi(\ln|\textrm{tg}^{1/b}(\frac{by}
{2})|)}{\sin^{1/2}(by)}}\frac{\psi(\ln|\textrm{tg}^{1/b}(\frac{by}{2})|)}
{\sin^{1/2}(by)}~dy= \langle \tilde{U}\psi|\tilde{U}\psi\rangle .
\end{eqnarray}
Thus, the mapping (\ref{interval}) is isometric and hence {\it
unitary}.

Now, let us see how the operator $\hat{w}$ transforms under the
unitary map (\ref{interval}). The transformation consists of the
change of an independent variable
\begin{equation}\label{indep}
    x\mapsto y:=\frac{2}{b}\textrm{arctan}(e^{bx}),
\end{equation}
which leads to
\begin{equation}\label{NEWoperator1}
   -\frac{ia}{2}\Big(2\cosh(bx)\frac{d}{dx}+b\sinh(bx)\Big)\mapsto
   -ia\frac{d}{dy}+i\frac{ab}{2}\cot(by),
\end{equation}
and re-scaling with respect to a dependent variable
\begin{equation}\label{NEWoperator2}
    -ia\frac{d}{dy}+i\frac{ab}{2}\cot(by)\mapsto\sin^{-1/2}(by)
\bigg(-ia\frac{d}{dy}+i\frac{ab}{2}\cot(by)\bigg)\sin^{1/2}(by)
=-ia\frac{d}{dy}.
\end{equation}
In the process of mapping
\begin{equation}\label{tran}
\hat{w}\mapsto \tilde{U}\,\hat{w}\,\tilde{U}^{-1} =
-ia\frac{d}{dy} =: \breve{w},
\end{equation}
we have used two identities: $\sin(by) = 1/\cosh(bx)$ and
$\sinh(bx)= - \cot(by)$.

Since $w>0$ (for $\textrm{O}>0$), we {\it assume} that the
spectrum of $\breve{w}$ consists of positive eigenvalues. To
implement this assumption, we  define
$\breve{w}:=\sqrt{\breve{w}^2}$ and consider the eigenvalue
problem
\begin{equation}\label{w^2}
- a^2\frac{d^2}{dy^2}f_{\nu}=\nu^2 f_{\nu},~~~~y\in (0,\pi/b).
\end{equation}
There are two independent solutions for each value of $\nu^2$
(where $\nu\in \dR$), namely: $\sin(\frac{\nu}{a}y)$ and
$\cos(\frac{\nu}{a}y)$. Removing this degeneracy leads to required
positive eigenvalues of $\breve{w}$. We achieve that in a standard
way by requiring that the eigenvectors  vanish at the boundaries,
i.e, at $y =0$ and $y= \pi/b$. As the result we get the following
spectrum
\begin{equation}\label{st}
f_{\nu}=N\sin(\frac{\nu}{a}y),~~~~\nu^2=(nab)^2,~~ n= 0, 1,2,\dots
\end{equation}
It should be noted that for $n=0$, the eigenvector is a null state
and thus the lowest eigenvalue is $\nu^2=(ab)^2$. Next, we define
the Hilbert space to be the closure of the span of the
eigenvectors  (\ref{st}). The operator $\breve{w}^2 = -
a^2\frac{d^2}{dy^2}$ is essentially self-adjoint on this span by
the construction. Due to the spectral theorem \cite{RaS} we may
define an essentially self-adjoint operator $\breve{w}=\sqrt{-
a^2\frac{d^2}{dy^2}}$ as follows
\begin{equation}\label{spec}
\breve{w}f_{\nu}:=\nu f_{\nu},~~~~\nu =ab,\,2ab,\,3ab,\dots
\end{equation}

We have considered the case  $w>0$. The case $w<0$ does not
require changing of the Hilbert space. The replacement
$\hat{w}\mapsto -\hat{w}$ leads to  $\nu \mapsto -\nu$.

Finally, we find that the inverse mapping from $L^2(\id,dy)$ to
$L^2(\dR,dx)$ for the eigenvectors of $\breve{w}$ yields
\begin{equation}\label{invmap}
    \sin\big(\frac{\nu}{a}y\big)=f_{\nu}(y)\mapsto
     \tilde{U}^{-1}f_{\nu}(y):={\psi}_{\nu}(x)=\frac{\sin\big(\frac{2\nu}
    {ab}\textrm{arctg}
    (e^{bx})\big)}{\cosh^{1/2}(bx)}.
\end{equation}

\subsection{Kasner-like case}

In the case (2a), the conditions $\textrm{O}_1 \textrm{O}_2 +
\textrm{O}_1\textrm{O}_3 + \textrm{O}_2 \textrm{O}_3 > 0$ with
$\textrm{O}_1<0, \textrm{O}_2
>0, \textrm{O}_3>0$ are satisfied in the following domains\footnote{The case (2b)
can be done by analogy.} for $\textrm{O}_k$
\begin{equation}\label{dom1}
    \textrm{O}_1 \in\, (-d_1,0),~~~\textrm{O}_2 \in\, (d_2,\infty),~~~\textrm{O}_3 \in\,
    (d_3,\infty),
\end{equation}
where $d_2 > d_1,$ and where $d_3 = d_1 d_2/ (d_2 - d_1)$ so $d_3
> d_1$. The full phase space sector of the Kasner-like evolution
is defined as the union
\begin{equation}\label{full}
\bigcup_{0<d_1<d_2}(-d_1,0)\times(d_2,\infty)\times(d_3,\infty)
\end{equation}

In the case of $\textrm{O}_2$ and $\textrm{O}_3$, the restrictions
for domains (\ref{dom1}) translate into the restrictions for the
corresponding domains for the observables $w_2$ and $w_3$, due to
(\ref{vi5}), and read
\begin{equation}\label{dom12}
    w_2 \in\, (D_2,\infty),~~~w_3 \in\,
    (D_3,\infty),
\end{equation}
where $D_2 = \kappa \gamma \lambda d_2$ and $D_3 = \kappa \gamma
\lambda d_3$. Thus, quantization of the $w_2$ and $w_3$
observables can be done by analogy to the Kasner-unlike case. The
spectra of the operators $\hat{w}_2$ and $\hat{w}_3$ are almost
the same as the spectrum defined by (\ref{spec}) with the only
difference that now $\nu> D_2$ and $\nu> D_3$,
respectively\footnote{Spectra are insensitive to unitary
transformations.}.

The case of $w_1$ requires special treatment.  Let us redefine the
elementary observables corresponding to the 1-st direction as
follows
\begin{equation}\label{red1}
    \Omega_1 := - \frac{O_1}{b \cosh(b\mathcal{A}_1)},~~~~\Omega_2:=
    \sinh(b\mathcal{A}_1).
\end{equation}
The transformation (\ref{red1}) is canonical, since
$\{\Omega_1,\Omega_2\}=1$, and invertible. The domains transform
as follows
\begin{equation}\label{r2}
O_1 \in (-d_1,0),~~~\mathcal{A}_1 \in \dR
~~~~~\longrightarrow~~~~~\Omega_1 \in (0,d_1/b)=:
(0,D_1),~~~\Omega_2 \in \dR.
\end{equation}
The observable $v_1$ in terms of redefined  observables reads
\begin{equation}\label{r3}
v_1 = \frac{ab}{\hbar}\,\Omega_1\,(1 + \Omega_2^2),~~~~v_1 \in
(0,\infty),
\end{equation}
where $ab/\hbar = 12\pi G\gamma\lambda$. To quantize observables
$\Omega_1$ and $\Omega_2$ we use the Schr\"{o}dinger
representation
\begin{equation}\label{r4}
\Omega_2 \rightarrow \hat{\Omega}_2f(x):=  -i\hbar\partial_x
f(x),~~~~~\Omega_1 \rightarrow \hat{\Omega}_1f(x):= x f(x),~~~~~f
\in L^2(0,D_1).
\end{equation}

Let us find an explicit form for the operator
$\,\frac{ab}{\hbar}(\widehat{\Omega}_1+\widehat{\Omega_1\Omega_2^2})$,
corresponding to  (\ref{r3}). Since $\Omega_1>0$, the following
classical equality holds
\begin{equation}\label{nov}
\Omega_1\Omega_2^2=\Omega_1^k\cdot\Omega_2\cdot\Omega_1^{1-k-m}\cdot
\Omega_2\cdot\Omega_1^{m},
\end{equation}
where $m, k\in \dR$. This may lead to many operator orderings at
the quantum level.  This issue is further discussed in the
appendix B and in the conclusion section.

We propose the following mapping  (we set $\hbar=1$)
\begin{equation}\label{novum}
\Omega_1\Omega_2^2 \rightarrow
\widehat{\Omega_1\Omega_2^2}:=\frac{1}{2}\bigg( \hat{\Omega}_1^k
\hat{\Omega}_2\,\hat{\Omega}_1^{1-k-m}\hat{\Omega}_2\,\hat{\Omega}_1^m
+ \hat{\Omega}_1^m\hat{\Omega}_2\,\hat{\Omega}_1^{1-k-m}
\hat{\Omega}_2\, \hat{\Omega}_1^k\bigg) =
-x\partial^2_{xx}-\partial_x+mkx^{-1},
\end{equation}
which formally ensures the symmetricity of
$\widehat{\Omega_1\Omega_2^2}$. The second equality in
(\ref{novum}) may be verified via direct calculations.

Now, we define the following unitary transformation $W$
\begin{equation}
L^2([0,D_1],dx)\ni f(x)\mapsto W f(x):=
\sqrt{\frac{y}{2}}f\bigg(\frac{y^2}{4}\bigg)\in
L^2([0,2\sqrt{D_1}],dy).
\end{equation}
One may verify that we have
\begin{eqnarray}
W\partial_x W^{\dag}=\frac{2}{y}\partial_y-\frac{1}{y^2},~~~~
W\partial^2_{xx}
W^{\dag}=\frac{4}{y^2}\partial^2_{yy}-\frac{8}{y^2}
\partial_y+\frac{5}{y^4}~.
\end{eqnarray}

Thus, the operator $W$ transforms (\ref{novum}) into
\begin{equation}
-\partial^2_{yy}+\frac{1}{y^2}\bigg(4mk-\frac{1}{4}\bigg).
\end{equation}

The eigenvalue problem for
$\widehat{\Omega}_1+\widehat{\Omega_1\Omega_2^2}$ reads
\begin{equation}\label{nov1}
\bigg(-\partial^2_{yy}+\frac{1}{y^2}\bigg(4mk-\frac{1}{4}\bigg)+
\frac{y^2}{4}\bigg)\Phi = \nu \,\Phi.
\end{equation}
Now, we can  see an  advantage of  the chosen ordering
prescription (\ref{novum}). It enables finding a very simple form
of the volume operator. Taking $k=m= 1/4$ turns (\ref{nov1}) into
\begin{equation}\label{w2}
\bigg(-\partial^2_{yy}+\frac{y^2}{4}-\nu\bigg)\Phi=0 .
\end{equation}
The problem is mathematically equivalent to the one dimensional
harmonic oscillator in a `box' with an edge equal to
$2\sqrt{D_1}$. There are two independent solutions for a given
$\nu$
\begin{eqnarray}
\Phi_{\nu,1}= N_1 e^{-y^2/{4}}~_1F_1\bigg(-\frac{1}{2}\nu+
\frac{1}{4},\frac{1}{2},\frac{y^2}{2}\bigg),\\
\Phi_{\nu,2}= N_2 ye^{-y^2/{4}}
~_1F_1\bigg(-\frac{1}{2}\nu+\frac{3}{4},\frac{3}{2},\frac{y^2}{2}\bigg),
\end{eqnarray}
where $\,_1F_1$ is a hypergeometric confluent function,
$\Phi_{\nu,1}$ and $\Phi_{\nu,2}$ are even and odd cylindrical
functions, respectively. A standard condition for the symmetricity
of the operator  defining the eigenvalue problem (\ref{w2}) leads
to the vanishing of  the wave functions at the boundaries (as the
box defines the entire size of the $1$-st direction). The solution
(after retrieving of $\hbar$ and $ab$) reads\footnote{We ignore
the solution $\Phi_{\nu,1}$ because it cannot vanish at $y=0$.}.
\begin{equation}\label{2w2}
\Phi=N ye^{-\frac{y^2}{4\hbar}}
~_1F_1\bigg(-\frac{1}{2}\frac{\nu}{ab}+\frac{3}{4},\frac{3}{2},
\frac{y^2}{2\hbar}\bigg).
\end{equation}
The solution (\ref{2w2}) vanishes at $y=0$ as $\Phi$  is an odd
function. The requirement of vanishing at $y=2\sqrt{D_1}$ leads to
the equation
\begin{equation}\label{w22}
    _1F_1\bigg(-\frac{1}{2}\frac{\nu}{ab\hbar}+\frac{3}{4},\frac{3}{2},
    \frac{2D_1}{\hbar}\bigg)=0.
\end{equation}
An explicit form of (\ref{w22}) reads
\begin{equation}\label{cond}
\sum_{n=0}^{\infty}\frac{\big(-\frac{1}{2}\frac{\nu}{ab}+\frac{3}{4}
\big)_n}{\big(\frac{3}{2}\big)_n}
\bigg(\frac{2D_1}{\hbar}\bigg)^n=0 ,
\end{equation}
where $(a)_n=a(a+1)\dots (a+n-1)$. It results from (\ref{cond})
that the eigenvalues must satisfy the condition: $\nu\geq ab$.

\subsection{Quantum volume operator}

Classically we have
\begin{equation}\label{vol 2}
V = |w_1 w_2 w_3|^{1/3} .
\end{equation}
One may verify that $v_k$ Poisson commute and $\hat{v}_k$ commute,
so we can take
\begin{equation}\label{vol3}
\widehat{V}^3 := \hat{v}_1 \hat{v}_2 \hat{v}_3 = |\hat{w}_1
\hat{w}_2 \hat{w}_3|.
\end{equation}
The eigenfunctions of the operator $\hat{w}_1 \hat{w}_2 \hat{w}_3$
have the form
$F^{\lambda_1,\lambda_2,\lambda_3}:={f_1}^{\lambda_1}(x_1){f_2}^{\lambda_2}(x_2)
{f_3}^{\lambda_3}(x_3)$, where ${f_i}^{\lambda_i}(x_i)$ is an
eigenvector of $\hat{w}_i$ with eigenvalue $\lambda_i$. The
closure of the span of $F^{\lambda_1,\lambda_2,\lambda_3}$ is a
Hilbert space, in which $\widehat{V}^3$ is a self-adjoint operator
(by construction).

Due to the spectral theorem on self-adjoint operators \cite{RaS},
we have
\begin{equation}\label{spec1}
 V = (V^3)^{1/3}~~~\longrightarrow~~~\widehat{V} F^{\lambda_1,\lambda_2,\lambda_3}
 := \square\,
 F^{\lambda_1,\lambda_2,\lambda_3},
\end{equation}
where
\begin{equation}\label{spec2}
 \square:= |\lambda_1 \lambda_2 \lambda_3|^{1/3}.
 \end{equation}

\subsubsection{Kasner-unlike case}

In the Kasner-unlike case we use the formula (\ref{spec}) to get
 \begin{equation}\label{spec3}
  \square =  |n_1n_2n_3|^{1/3}\,ab,~~~~~n_1, n_2, n_3 \in
  1,2,3,\dots ,
 \end{equation}
which shows that the spectrum of the volume operator does not have
equally distant levels.  The volume $\square$ equal to {\it zero}
is not in the spectrum. There exist a {\it quantum} of the volume
which equals $\triangle:=ab = 12\pi G\gamma\lambda\hbar$, and
which defines the lowest value in the spectrum.

\subsubsection{Kasner-like case}

The spectrum in this case reads
\begin{equation}
  \square :=  \bigcup_{0<d_1<d_2}\square_{d_1,d_2},~~~~\square_{d_1,d_2}:= \{\lambda_{d_1}
     \lambda_{d_2}
    \lambda_{d_3}\,|\, d_3 = d_1 d_2/ (d_2 - d_1)\},
\end{equation}
where $\lambda_{d_1}$ is any value subject to the condition
(\ref{cond}), $\lambda_{d_2}>D_2$ and $\lambda_{d_3}>D_3$ are
given by (\ref{spec}). The volume $\square$ equal to {\it zero} is
not in the spectrum.

\subsection{Evolution}

In this section we ignore the restrictions concerning the domains
of $\textrm{O}_1$, $\textrm{O}_2$ and $\textrm{O}_3$, and we
assume that the Hilbert space of the system is
$L^2(\mathbb{R}^3,dxdydz)$. An  inclusion of the restrictions
would complicate the calculations without bringing any qualitative
change into the picture of evolution.

The generator of evolution determined in (\ref{Htrue}) may be
formally quantized, due to (\ref{osy}), as follows
\begin{equation}
\mathbb{H}\mapsto \hat{\mathbb{H}}
=-\hbar^2(\partial_y\partial_z+\partial_x\partial_z+\partial_x\partial_y).
\end{equation}
Since it is self-adjoint in $L^2(\mathbb{R}^3,dxdydz)$, a quantum
evolution can be defined by an unitary operator
\begin{equation}
 U= e^{-i\hbar\tau(\partial_y\partial_z+\partial_x\partial_z+\partial_x\partial_y)},
 ~~~~~\tau \in \dR.
\end{equation}
Let us study  an evolution of the expectation value of the
directional volume $\hat{v}_1$
\begin{equation}
 \langle\psi |U^{-1}\hat{v}_1U | \psi\rangle
\end{equation}
Since $\hat{v}_1$ does not depend on $y$ and $z$, we simplify our
considerations by taking
\begin{equation}
 U_1= e^{-i\hbar\tau(\partial_z+\partial_y)\partial_x}.
\end{equation}
If we are interested in the action of $U_1$ on the functions
$f(x)\in L^2(\mathbb{R},dx)$, then the derivatives
$-i\frac{d}{dy}$ and $-i\frac{d}{dz}$ occurring in $U_1$ commute
and, being self-adjoint, lead finally to real numbers. Let us call
them $k_y$ and $k_z$, respectively, and let us introduce the
parameter $k=k_y+k_z$. Hence, $U_1$ further simplifies and reads
\begin{equation}
 U_1= e^{k\hbar\tau\partial_x}.
\end{equation}
The action of $U_1$ on $f(x)$ reads
\begin{equation}
 U_1f(x)=f(x+k\hbar\tau) .
\end{equation}
We recall that under the unitary mapping
$L^2(\mathbb{R},dx)\mapsto L^2(\mathbb{I},dy)$, defined by
(\ref{interval}), the operator $\hat{v}_1$ becomes
$-ia\frac{d}{dy}$ on  $L^2(\mathbb{I},dy)$. Now, let us study an
action of operator $U_1$ on the functions $\varphi(y)\in
L^2(\mathbb{I},dy)$. Straightforward calculation leads to
\begin{equation}
L^2(\mathbb{I},y)\ni\varphi(y)\mapsto\frac{\varphi(\frac{2}{b}\arctan(e^{bx}))}
{\cosh^{1/2}(bx)}\in L^2(\mathbb{R},x),
\end{equation}
and we have
\begin{equation}
 U_1\frac{\varphi(\frac{2}{b}\arctan(e^{bx}))}{\cosh^{1/2}(bx)}=
 \frac{\varphi(\frac{2}{b}\arctan(e^{bx+bk\hbar\tau}))}{\cosh^{1/2}(bx+bk\hbar\tau)}
\end{equation}
The transformation $\tilde{U}^{-1}$ gives
\begin{equation}\label{evol}
 \frac{\varphi(\frac{2}{b}\arctan(e^{bx+bk\hbar\tau}))}{\cosh^{1/2}(bx+bk\hbar\tau)}
 \mapsto\frac{\varphi(\frac{2}{b}\arctan(e^{bk\hbar\tau}\tan(\frac{by}{2})))}
 {\sqrt{\frac{1}{2}\sin(by)(\tan(\frac{by}{2})e^{bk\hbar\tau}+\cot(\frac{by}{2})
 e^{-bk\hbar\tau})}}=:\varphi_\tau(y),
\end{equation}
where $\varphi_{\tau=0}(y)=\varphi(y)$. Now, we observe that the
symmetricity condition
\begin{equation}
    \langle\varphi_{\tau}(y)|\hat{v}_1\varphi_{\tau}(y)\rangle = \langle\hat{v}_1
    \varphi_{\tau}(y)|\varphi_{\tau}(y)\rangle
\end{equation}
leads to
\begin{equation}\label{evol2}
    \overline{\varphi}_{\tau}(\frac{\pi}{b})\varphi_{\tau}(\frac{\pi}{b})-
    \overline{\varphi}_{\tau}(0)\varphi_{\tau}(0)=0.
\end{equation}
We use the result (\ref{evol}) to calculate the limits
\begin{equation}
    \lim_{y\longrightarrow 0}\varphi_{\tau}(y)=e^{\frac{bk\hbar\tau}{2}}
    \varphi_{0}(0),~~~~
    \lim_{y\longrightarrow \frac{\pi}{b}}\varphi_{\tau}(y)=e^{-\frac{bk\hbar\tau}
    {2}}\varphi_{0}(\frac{\pi}{b}),
\end{equation}
which turns (\ref{evol2}) into
\begin{equation}\label{any}
    \overline{\varphi}_0(\frac{\pi}{b})\varphi_0(\frac{\pi}{b})e^{-bk\hbar\tau}-
    \overline{\varphi}_0(0)\varphi_0(0)e^{bk\hbar\tau}=0.
\end{equation}
It is clear that (\ref{any}) can be satisfied   $\,\forall \tau$
{\it iff} $\,{\varphi}_0(\frac{\pi}{b})=0=\varphi_0(0)$. States
with such a property belong to the domain of $\breve{w}$  defined
by (\ref{spec}).

In order to construct the `evolving states' that vanish at the
boundaries, consider the basis vectors $f_n(y)=e^{i2bny}$. Then,
$f_n(y)-f_m(y)$ satisfy the condition (\ref{any}). Making use of
(\ref{evol}) we get
\begin{equation}
    f_n(y,\tau)=\bigg(\frac{i-e^{bk\hbar\tau}\tan(\frac{by}{2})}
    {i+e^{bk\hbar\tau}
    \tan(\frac{by}{2})}\bigg)^{2n}
    \sqrt{\frac{1+\tan^2(\frac{by}{2})}{e^{-bk\hbar\tau}+
    e^{bk\hbar\tau}\tan^2(\frac{by}{2})}},
\end{equation}
where $f_n(y,\tau):= f_{n,\tau}(y)$. Moreover we have
\begin{equation}
    -ia\frac{d}{dy}f_n(y,\tau)=-i\frac{ab}{2}(1+\tan^2(\frac{by}{2}))f_n(y,\tau)
    \frac{1}{1+e^{2bk\hbar\tau}\tan^2(\frac{by}{2})}
    \bigg(\frac{(1-e^{2bk\hbar\tau})\tan(\frac{by}{2})}{1+\tan^2(\frac{by}{2})}+
    i4ne^{bk\hbar\tau}\bigg).
\end{equation}
Using the substitution $x=\tan(\frac{by}{2})$ we get
\begin{eqnarray}\nonumber
    \langle f_m|-ia\frac{d}{dy}f_n\rangle=\\
    -ia\int_0^{\infty}
    \bigg(\frac{i-e^{bk\hbar\tau}x}{i+e^{bk\hbar\tau}x}\bigg)^{2(n-m)}
    \frac{(e^{-bk\hbar\tau}-e^{bk\hbar\tau})x}{(e^{-bk\hbar\tau}+e^{bk\hbar\tau}x^2)^2}
    ~dx\\ \nonumber
    +4an\int_0^{\infty}
    \bigg(\frac{i-e^{bk\hbar\tau}x}{i+e^{bk\hbar\tau}x}\bigg)^{2(n-m)}
    \frac{1+x^2}{(e^{-bk\hbar\tau}+e^{bk\hbar\tau}x^2)^2}~dx .
\end{eqnarray}
Another substitution $z=e^{bk\hbar\tau}x$ leads to
\begin{eqnarray}\nonumber
    \langle f_m|-ia\frac{d}{dy}f_n\rangle=\\
    -ia(e^{-bk\hbar\tau}-e^{bk\hbar\tau})\int_0^{\infty}
    \bigg(\frac{i-z}{i+z}\bigg)^{2(n-m)}\frac{z}{(1+z^2)^2}
    ~dz\\ \nonumber
    +4an\int_0^{\infty}
    \bigg(\frac{i-z}{i+z}\bigg)^{2(n-m)}\frac{e^{bk\hbar\tau}+
    e^{-bk\hbar\tau}z^2}{(1+z^2)^2}
    ~dz
\end{eqnarray}
Finally, we obtain
\begin{equation}
 \langle f_m|-ia\frac{d}{dy}f_n\rangle= \left\{
\begin{array}{l}
  \frac{ia}{4(n-m)^2-1}(1-8n(n-m))\sinh(bk\hbar\tau),~~n\neq m \\
  ia\sinh(bk\hbar\tau)+2\pi na\cosh(bk\hbar\tau),~~n= m .\\
\end{array}\right.
\end{equation}
Now, let us introduce
$g_{nm}(y,\tau):=\frac{f_n(y,\tau)-f_m(y,\tau)}{\sqrt{\frac{2\pi}{b}}}$
so that $\|g_{nm}\|=1$. One has
\begin{equation}\label{expec}
 \langle g_{nm}|-ia\frac{d}{dy}g_{nm}\rangle=(n+m)ab\cosh(bk\hbar\tau)=\frac{n+m}{2}
 \Delta\cosh(bk\hbar\tau).
\end{equation}

The expectation value of the operator (\ref{expec}), defining the
volume operator, is similar to the classical form (\ref{vi5}). The
vectors $g_{nm}$ may be used in the construction of a basis of the
space of states such that ${\varphi}_0(\frac{\pi}{b})= 0
=\varphi_0(0)$.

\section{Summary and conclusions}
Turning the Big Bang into the Big Bounce  in the Bianchi I
universe is due to the modification of the model at the {\it
classical} level by making use of the loop geometry (in complete
analogy to the FRW case). The modification is parameterized by a
continuous parameter $\lambda $ to be determined from
observational astro-cosmo data.

The reduced phase space of our system is higher dimensional with
nontrivial boundaries. This requires introducing  a few new
elements, comparing to the FRW case, into our method: 1.  Our
parameter does {\it not} need to coincide with the scalar field,
commonly used in loop quantum cosmology, and it simplifies the
form of a volume operator. 2. We introduce the so-called {\it
true} Hamiltonian. It generates a flow in the family of volume
quantities, enumerated by the evolution parameter. It proves  an
independence of the spectrum of the volume operator on the
evolution. 3. We divide the phase space of the system into two
distinct regions: Kasner-like and Kasner-unlike. 4. We identify
domains, spectra and eigenvectors of \emph{self-adjoint}
directional volumes and total volume operators in the
Kasner-unlike case. 5. We identify the {\it peculiarity} of the
Kasner-like case due to complicated boundary of the phase space
region. We propose to overcome this problem by dividing this
region further into smaller regions, but with simpler boundaries.
6. Given a small subregion for Kasner-like case, we propose a
canonical {\it redefinition} of phase space coordinates in such a
way, that we can arrive at relatively simple form of volume
operator and at the same time can simply encode the boundary of
the region into the Schr\"odinger representation. Then, from a
number of different operator orderings we chose the simplest one.
We find domain, spectrum and eigenvectors of the volume operator.
The spectrum is given in an implicit form in terms of special
functions. 7. Having the true (self-adjoint) Hamiltonian, we
introduce an {\it unitary} operator with an evolution parameter.

The spectrum of the volume operator, parameterized by $\lambda$,
is bounded from below and {\it discrete}.  An evolution of the
expectation value of the volume operator is similar to the
classical case. We have presented the evolution of only a single
directional volume operator. One may try to generalize this
procedure to the {\it total} volume operator. Analyzes, in the
case of the Kasner-like dynamics, are complicated and will be
presented elsewhere.

Discreteness of space at the quantum level may lead to a {\it
foamy} structure of spacetime at the semi-classical level.  A
possible astro-cosmo observation of {\it dispersion} of photons
travelling over cosmological distances across the Universe might
be used to determine the value of otherwise free parameter
$\lambda $. The discreteness is also specific to the FRW case
\cite{Malkiewicz:2009qv}. The difference is that in the Bianchi I
case the variety of possible quanta of a volume is much richer. On
the other hand, the Bianchi type cosmology seems to be more
realistic than the FRW case, near the cosmological singularity.
Thus, an expected foamy structure of space may better fit
cosmological data. Various forms of discreteness of space may
underly many approaches in fundamental physics. So its examination
may be valuable.

Quantum cosmology calculations are plagued by  quantization {\it
ambiguities}. In particular, there exists a huge freedom in
ordering of elementary operators defining compound observables,
which may lead to different quantum operators. Classical
commutativity of  variables does not extend to corresponding
quantum operators. Such ambiguities can be largely reduced after
some quantum astro-cosmo data become available. Confrontation of
theoretical predictions against these data would enable finding
realistic quantum cosmology models.

Our nonstandard loop quantum cosmology method, successfully
applied so far mainly to the FRW type models, seems to be  highly
efficient and deserves  further development and application to
sophisticated cosmological models of general relativity.


\appendix

\section{Unitarily non-equivalent volume operators}

In both Kasner-like and Kasner-unlike cases, we have reduced the
Hilbert space by removing the double {\it degeneracy} of
eigenvalues for the volume operators (see the discussion after
equations (\ref{w^2}) and (\ref{w2})). We have used the `natural'
condition that the wave function should vanish at the boundaries
of an interval.  However, there are also other mathematically
well-defined choices for the boundary conditions. We will
demonstrate this non-uniqueness for the Kasner-unlike case.
Similar reasoning applies to another case.

Let us begin with the equation (\ref{w^2})
\begin{equation}
-a^2\frac{d^2}{dy^2}f=\nu^2f,~~~~y\in (0,\pi/b),
\end{equation}
which has the solution
\begin{equation}
f_{\nu}=N_1\sin(\frac{\nu}{a}y)+N_2\cos(\frac{\nu}{a}y),~~~~N_1,N_2\in
\mathbb{C},
\end{equation}
for each value of $\nu\in \dR_+$ (~$\nu\mapsto -\nu$ does not
produce any new space of solutions). Our task is the determination
of self-adjointness  of $\breve{w}:=\sqrt{-a^2\frac{d^2}{dy^2}}$
and removing  the double degeneracy of eigenvalues. The
symmetricity condition reads
\begin{equation}\label{sym}
    \int_{I}\bar{f}f''=\bar{f}f'\bigg|^{\pi/b}_{0}-\bar{f}'f\bigg|^{\pi/b}_0
    +\int_I\bar{f}''f.
\end{equation}

We can set:
\begin{itemize}
    \item  $f(0)=f(\pi/b)=0 \Rightarrow
    f_{\nu}=\sin(\frac{\nu}{a}y),~~\nu=ab,2ab,3ab,\dots$
    \item $f'(0)=f'(\pi/b)=0 \Rightarrow
    f_{\nu}=\cos(\frac{\nu}{a}y),~~\nu=0,ab,2ab,3ab,\dots$
    \item $f(0)=f'(\pi/b)=0 \Rightarrow
    f_{\nu}=\sin(\frac{\nu}{a}y),~~\nu=\frac{1}{2}ab,\frac{3}{2}ab,
    \frac{5}{2}ab,\dots$
\end{itemize}
where $ab = 12\pi\hbar G\gamma\lambda.$ All these choices are
non-equivalent, since they lead to different spectra.

\section{Standard quantization}

Let us change the coordinates of the Kasner-like sector phase
space $(\Omega_1,\Omega_2)$, defined by (\ref{red1}), into a new
canonical pair as follows
\begin{equation}
X:=\sqrt{2\Omega_1}~~\textrm{and}~~P:=\Omega_2\sqrt{2\Omega_1},
\end{equation}
where
\begin{equation}
(X,P)\in (0,\sqrt{2d_1/b})\times\dR,~~~~\{X,P\}=1.
\end{equation}
In the new variables the volume (\ref{r3}) reads
\begin{equation}
\frac{1}{4\pi G\gamma\lambda}~v_1=\frac{1}{2}P^2+ \frac{1}{2} X^2.
\end{equation}
Thus, in these  variables  the volume has a form of the
Hamiltonian of the harmonic oscillator in a `box'
$(0,\sqrt{2d_1/b})$.

In the Schr\"odinger representation, i.e. $\hat{X}:=x$ and
$\hat{P}:=-i\hbar\partial_{x}$, a standard quantization yields
\begin{equation}
    \frac{1}{4\pi G\gamma\lambda}~\hat{v}=-\frac{\hbar^2}{2}\partial^2_{xx}
    + \frac{1}{2}x^2 ,
\end{equation}
which corresponds to the `nonstandard' quantization (\ref{nov1})
with the parameters $m=k=1/4$ and $y = \sqrt{2} x$ (with $\hbar
=1$).

Thus, we can see that the prescription defined by (\ref{nov}) and
(\ref{novum}) includes not only a standard prescription, but many
others.  We have completed only one, corresponding to the well
known harmonic oscillator, as an illustration.

\end{document}